\documentclass[prl,twocolumn,superscriptaddress,amsmath]{revtex4-1}

\usepackage{natbib}
\usepackage{graphicx}
\usepackage{braket}
\usepackage[usenames,dvipsnames]{color}
\usepackage[FIGTOPCAP,raggedright,nooneline,bf,footnotesize]{subfigure}
\usepackage{hyperref}
\begin{document}

\title{Inherent polarization entanglement generated from a monolithic semiconductor chip}

\author{Rolf T. Horn$^*$}
\affiliation{Institute for Quantum Computing and Department of Physics and Astronomy,
University of Waterloo, 200 University Avenue W, Waterloo, Ontario, N2L 3G1, Canada}

\author{Piotr Kolenderski}
\affiliation{Institute for Quantum Computing and Department of Physics and Astronomy,
University of Waterloo, 200 University Avenue W, Waterloo, Ontario, N2L 3G1, Canada}
\affiliation{Institute of Physics, Faculty of Physics, Astronomy and Informatics, Nicolaus Copernicus University, Grudziadzka 5, 87-100 Torun, Poland}

\author{Dongpeng Kang} \author{Payam Abolghasem}
\affiliation{The Edward S. Rogers Sr. Department of Electrical and Computer Engineering, University of Toronto, 10 King's College Road, Toronto, Ontario M5S 3G4, Canada}

\author{Carmelo Scarcella}
\author{Adriano Della Frera}
\author{Alberto Tosi}
\affiliation{Politecnico di Milano, Dipartimento di Elettronica, Informazione e Bioingegneria,Piazza Leonardo da Vinci 32, 20133 Milano, Italy}

\author{Lukas G. Helt}
\affiliation{Department of Physics, University of Toronto, 60 St. George St., Toronto, Ontario M5S 1A7, Canada}
\affiliation{Centre for Ultrahigh bandwidth Devices for Optical Systems (CUDOS), MQ Photonics Research Centre, Department of Physics and Astronomy, Faculty of Science, Macquarie University, NSW 2109, Australia}
\author{Sergei V. Zhukovsky}
\affiliation{Department of Physics, University of Toronto, 60 St. George St., Toronto, Ontario M5S 1A7, Canada}
\affiliation{DTU Fotonik - Department of Photonics Engineering, Technical University of Denmark, Oersteds Pl. 343, DK-2800 Kgs. Lyngby, Denmark}
\author {John E. Sipe}
\affiliation{Department of Physics, University of Toronto, 60 St. George St., Toronto, Ontario M5S 1A7, Canada}

\author{Gregor Weihs}
\affiliation{Institute for Experimental Physics, University of Innsbruck, Technikerstrasse 25, 6020 Innsbruck, Austria}

\author{Amr S. Helmy}
\affiliation{The Edward S. Rogers Sr. Department of Electrical and Computer Engineering, University of Toronto, 10 King's College Road, Toronto, Ontario M5S 3G4, Canada}

\author{Thomas Jennewein}
\affiliation{Institute for Quantum Computing and Department of Physics and Astronomy,
University of Waterloo, 200 University Avenue W, Waterloo, Ontario, N2L 3G1, Canada}

\date{\today}

\begin{abstract}
\textbf{Creating miniature chip scale implementations of optical quantum information protocols is a dream for many in the quantum optics community.  This is largely because of the promise of stability and scalability.  Here we present a monolithically integratable chip architecture upon which is built a photonic device primitive called a Bragg reflection waveguide (BRW). Implemented in gallium arsenide, we show that, via the process of spontaneous parametric down conversion, the BRW is capable of directly producing polarization entangled photons without additional path difference compensation, spectral filtering  or post-selection. After splitting the twin-photons immediately after they emerge from the chip, we perform a variety of correlation tests on the photon pairs and show non-classical behaviour in their polarization.  Combined with the BRW's versatile architecture our results signify the BRW design as a serious contender on which to build large scale implementations of optical quantum processing devices.}
\end{abstract}
\maketitle

The use of non-linear optical effects has lead to an era of high-level quantum experiments, where quantum entanglement between pairs of photons can be achieved with very high count rates, quality and flexibility.  It has relied heavily on a photon pair producing process called spontaneous parametric down conversion (SPDC)\cite{Louisell_1961.art,Klyshko_1967.art}, where under conservation of energy and momentum, a ``pump'' photon spontaneously decays, leaving a pair of daughter photons in its place.  SPDC is now easily achieved using lasers and the same materials and principles involved in the more classical processes of sum frequency and second harmonic generation.

More challenging has been the creation of \emph{entangled} photon pairs in a degree of freedom, in our case polarization, that is uncorrelated with other aspects of the pair producing process \cite{Kim2002}. The challenge is indirectly tied to the very same principles and materials that facilitate the creation of photon pairs in the first place.  Birefringence, which makes phase matching the pump and daughter photons an easy task, also tends to provide distinguishing information that hinders the production of entanglement.  Despite this natural limitation, bulk-crystal photon sources have created entanglement via clever interference techniques, and often include additional compensating optics \cite{Ueno2012}.  While some outstanding results have been achieved \cite{Yao_2012.art}, too often the interferometers and compensation procedures are unstable, requiring daily maintenance, strict environmental conditions, or complicated automation.   It is generally thought that, to achieve larger scale demonstrations of optical quantum information protocols, the techniques developed to date should be integrated in some sense \cite{O'Brien_2007,Politi_2008} -- a so called optical bench on a chip.

\begin{figure*}[t]
\includegraphics[width=1.99\columnwidth]{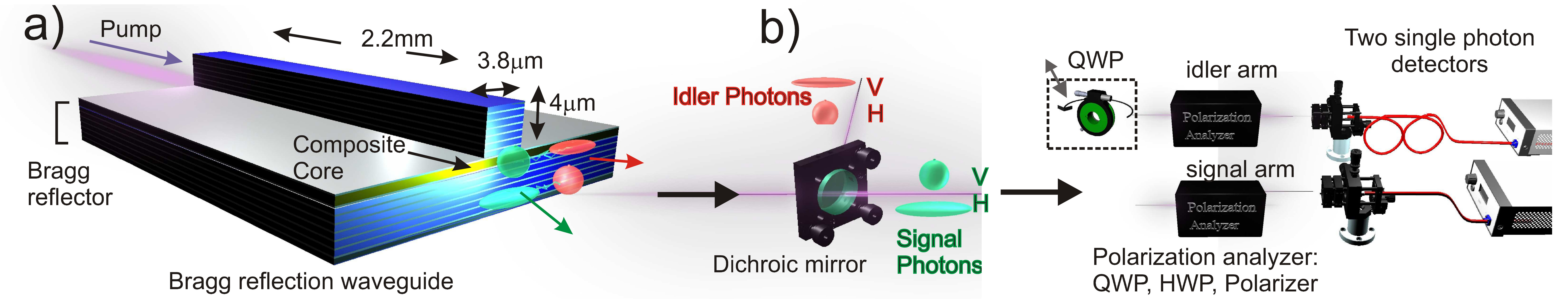}
\caption {The AlGaAs Bragg Reflection Waveguide and the experimental setup. (a) Layers of varying composition of aluminum gallium arsenide act as Bragg reflectors and sandwich a core layer (yellow).  A pump photon (purple) entering the BRW can spontaneously decay into a signal photon (green) and an idler photon (red) with either of two polarizations (H - elongated, V - rounded) via SPDC.  The characteristics of generated photon pairs can be tuned by altering the ridge dimensions and the epitaxial growth parameters.  (b) Photon pairs created in the BRW were split according to their color into shorter wavelength (signal) and longer wavelength (idler) arms.  Each photon's polarization was measured in various reference frames using a quarter-wave plate (QWP) and a half-wave plate (HWP), and a polarizer.  The phase between H(V) and V(H) photon pairs was modulated by inserting and tilting an additional QWP in only the idler arm. }
\label{fig:Figure1}
\end{figure*}

Semiconductors are an obvious platform for integration, but the material physics are vastly different from what is found in traditional SPDC based resources of quantum information. Recently publicized as a strong source of photon pairs \cite{Horn_2012.art} the gallium arsenide (GaAs) based Bragg reflection waveguide (BRW) was shown to have a distinct advantage over other semiconductor sources, due largely to its monolithic architecture -- its layered design underpinning many photonic devices. Using the same techniques as employed in bulk-crystal sources \cite{Valles2013}, it has recently been shown that the BRW can deliver polarization entangled photons through the use of post-selection. Here we show a significant advantage by demonstrating the intrinsic capability of the BRW to directly produce polarization entangled photon pairs, without any additional interferometry, spectral filtering, compensation or post-selection.
Combined with the fact that the structure can be its own pump laser \cite{Bijlani_11.art}, our results set the stage for the BRW to be a self-contained room temperature resource of entanglement occupying no more than a few square millimeters of chip real-estate.

Built on a GaAs substrate, the BRW consists entirely of layers of GaAs containing different amounts of aluminum. A subset of the layers function as Bragg reflectors, one lying below and one on top of a core region where light can be guided. A schematic is shown in Fig.~\ref{fig:Figure1}(a). The Bragg reflectors distinguish the waveguide. In addition to optical modes guided via the process of total internal reflection (TIR), the reflectors confine modes of light to the core region through interference.  Aptly named, the dispersion characteristics of Bragg modes are mostly independent from those of the more traditional TIR modes.   This degree of freedom allows the designer to use Bragg interference to create structures where the daughter TIR modes of SPDC are perfectly phase matched to the pump Bragg modes \cite{Helmy_06.art} -- a technique called modal phase matching.

Modal phase matching based on Bragg reflection has important ramifications for integrated optical quantum devices that employ SPDC. Not only does the BRW allow for the production of photon pairs in non-birefringent semiconductors like GaAs \cite{Horn_2012.art}, but the detrimental effects of any residual birefringence on the production of polarization entangled photons are minimal and design related. In fact, the GaAs based BRW can be engineered such that the optical properties of orthogonally polarized TIR modes are identical.  Specifically, inherent polarization entanglement becomes possible for frequency \emph{non-degenerate} co-linear type-II modal phase matching where daughter photons are generated at different frequencies but guaranteed to have orthogonal polarization.  Because there is little to no birefringence, the daughter photons can emerge orthogonally polarized via two different decay ``channels'': In one channel, there is a probability amplitude $A_{HV}(\omega_1,\omega_2)$ where the higher (lower) frequency photon is horizontally (vertically) polarized. Alternatively, there is a probability amplitude $A_{VH}(\omega_1,\omega_2)$ where the higher (lower) frequency photon is vertically (horizontally) polarized. There is no need for compensation of any kind, and since the photons within a pair always have different frequencies, each pair can be easily split into distinct spatial arms, which we call signal and idler, by using a dichroic beam splitter (DBS). The expected state is polarization entangled and written as:
\begin{multline}
\ket{\Psi} = \int d\omega_1 d\omega_2\Bigl[A_{HV}(\omega_1,\omega_2)\ket{\omega_1,H}_s\ket{\omega_2,V}_i+\Bigr. \\\Bigl.A_{VH}(\omega_1,\omega_2)e^{i\gamma}\ket{\omega_2,V}_s\ket{\omega_1,H}_i\Bigr],
\label{eq:state:BRW:dichroic}
\end{multline}
where H(V) symbolizes horizontal(vertical) polarization, $\omega_1$ and $\omega_2$  represent the angular frequency of the photon, and $\gamma$ accounts for a relative phase which can be controlled by placing and tilting a birefringent element such as a quarter wave plate (QWP) in one of either the signal or idler arm. Note that the sum, $A_{HV}(\omega_1,\omega_2)+A_{VH}(\omega_1,\omega_2)$, is proportional to the joint spectral amplitude of the photon pairs generated by the BRW, and that the index interchange in the second term of Eq.~\ref{eq:state:BRW:dichroic} arises from the DBS transformation, which collapses the entire spectral output of the BRW to either the signal or idler arm. For a more in depth analysis of the generated quantum states in the presence of a DBS, please refer to the Supplementary Information.

\begin{figure}[t]
\centering
\begin{tabular}{c c}
\subfigure[H-polarized]{\includegraphics[width=0.49\columnwidth]{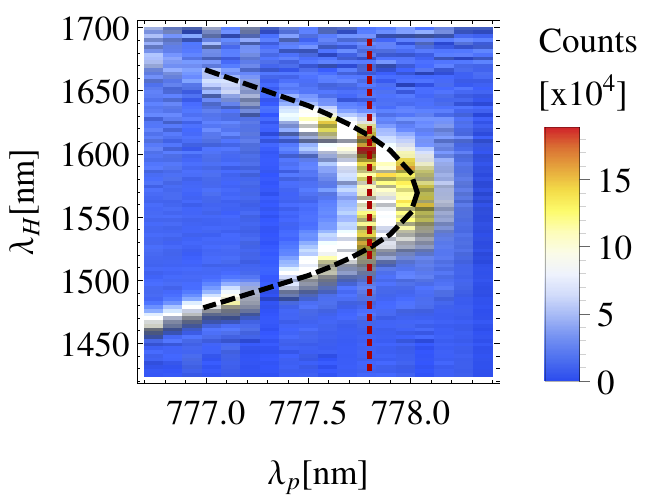}} &
\subfigure[V-polarized]{\includegraphics[width=0.49\columnwidth]{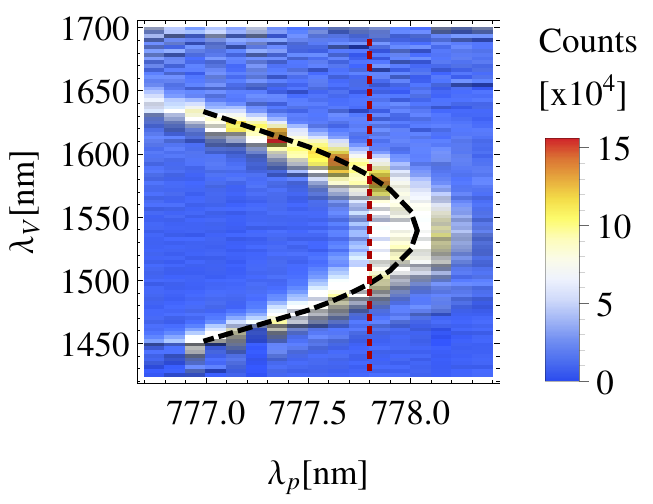}}
\end{tabular}
\caption{Intensity plots of the measured single photon spectra for (a) horizontal and (b) vertical polarized SPDC-output from the chip, versus the wavelength of the pump laser. The dashed lines are theoretical predictions indicating where the TIR modes are perfectly phase matched to a single frequency pump. The CW pump wavelength used for the polarization entanglement measurements is marked with the red dotted lines. }
\label{fig:Figure3}
\end{figure}

Two-photon polarization entanglement quality is inexorably linked to the absence of any information that might determine the polarization of either photon before it is measured. In our case, it is pre-dominantly limited by the amount of spectral overlap between $A_{HV}(\omega_1,\omega_2)$ and $A_{VH}(\omega_2,\omega_1)$. For this reason we measured the spectrum of both H and V polarized photons produced by the BRW for a range of pump wavelengths ($\lambda_p = 776 - 778$~nm). Details of this measurement can be found in the Methods section. The results, depicted in Fig.~\ref{fig:Figure3}, show clear similarities between the two orthogonally polarized spectra.  Importantly, they point to an inherent ambiguity in determining the polarization of any one photon by acquiring knowledge of its wavelength. Our observations imply that, for our experimental setup, a pump frequency setting of $\lambda_p\approx777.9$~nm is optimal for producing frequency non-degenerate polarization entangled photons.

It is worth emphasizing that such spectral evidence of two different decay channels is in contradistinction to the spectra of typical type-II SPDC where, when pumping away from degeneracy, only a single decay channel is observed. In those sources, the underlying crystal lattice asymmetry yields significant birefringence that prevents a second decay channel from being simultaneously phase matched.  Here however, the absence of lattice \emph{asymmetry} in GaAs (no birefringence) means that fundamentally, where an H-V signal-idler pair can have the correct phase matching, so can a V-H pair. From a polarization entanglement perspective, this is an extremely desirable phase matching scenario.  The BRW design is therefore advantageous because it not only solves the historically challenging problem of phase matching in non birefringent media, but the possibility of entanglement emerges as a naturally occurring byproduct.

In order to test the capabilities of the BRW to produce frequency non-degenerate polarization entanglement we subsequently performed two standard tests. The first measured entanglement visibilities, verifying that we indeed observe a coherence between the photon pairs. The second involved a detailed examination of the polarization state by way of quantum state tomography. Details are provided in the Methods section.

\begin{figure}[t]
\centering
\begin{tabular}{c}
\subfigure[\ H/V basis]{\includegraphics[width=0.9\columnwidth]{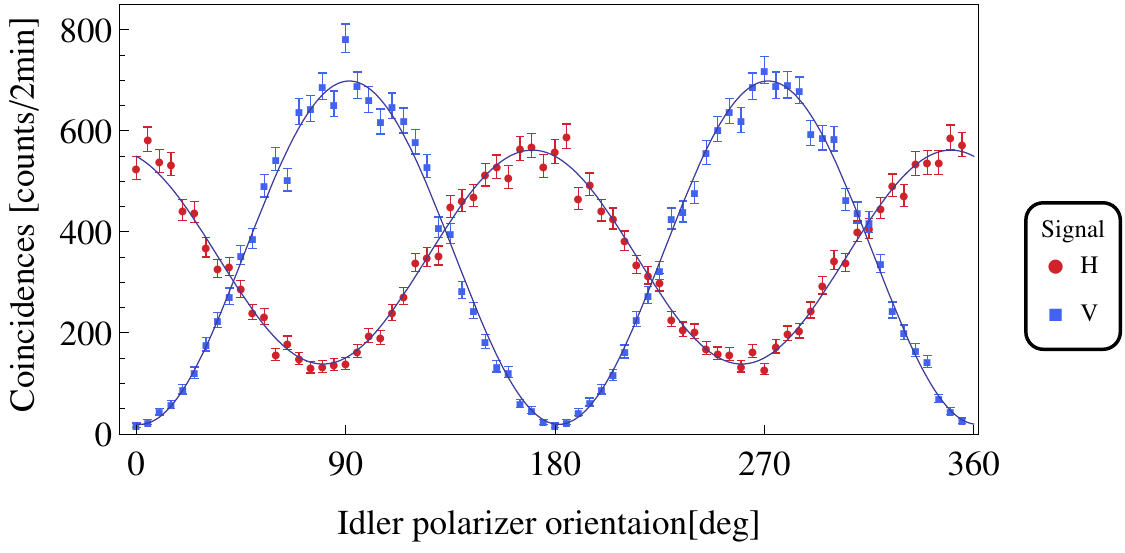} }\\
\subfigure[\ D/A basis]{\includegraphics[width=0.9\columnwidth]{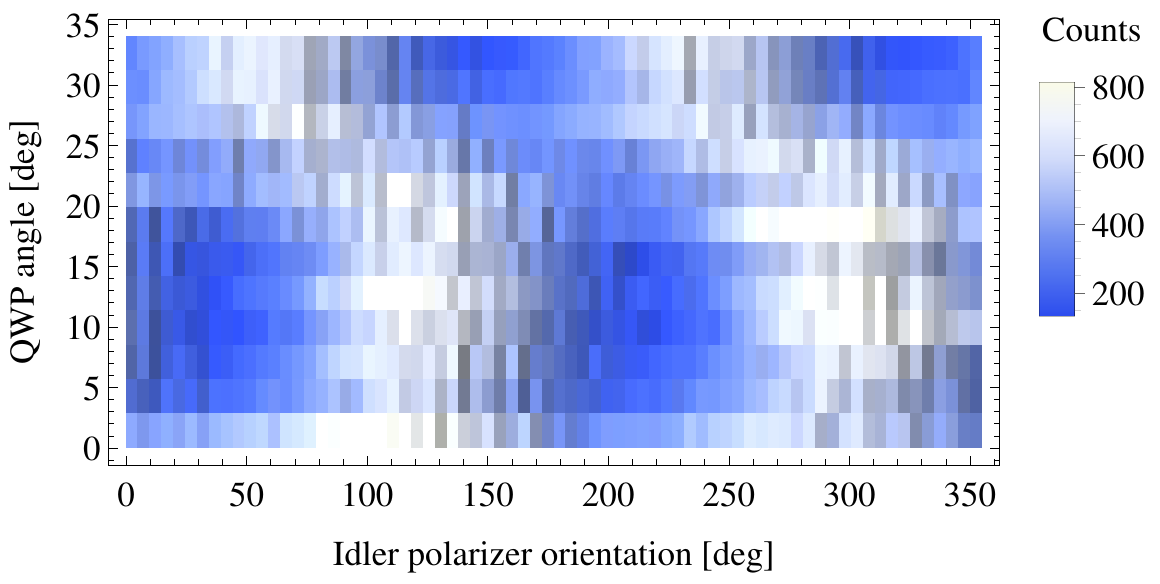}}\\
\end{tabular}
\caption{Polarization entanglement analysis. Panel (a) shows the pair count rates when the signal polarization analyzer measured horizontal (circles, red online) or vertical (squares, blue online) polarization while the idler polarization analyzer made many linear polarization measurements. For the diagram, 0$^\circ$ corresponds to a measurement of vertical polarization.  Without background subtraction, visibilities for the H,V  basis choices were $64\pm 3\%$, $96\pm 3\%$, respectively. Plot (b) shows data taken for the signal polarization analyzer set to 45$^{\circ}$.  A phase $\gamma$ was controlled by inserting and tilting a QWP in the idler arm. For the different tilt angles the highest visibility was $67\pm 3 \% $.}
\label{fig:Figure4}
\end{figure}

Entanglement visibility measurements illuminate the type of nonclassical polarization correlations that are expected. Pair count rates were recorded as follows:  A basis or reference frame was chosen by the polarization analyzer in the signal arm while the idler arm made many polarization measurements in continuous fashion. For a polarization entangled state, pair count rates are expected to behave sinusoidally in more than one basis as the idler arm's polarization analyzer changes.  For the ``HV'' basis, visibilities are shown in Fig.~\ref{fig:Figure4}(a).  Fig.~\ref{fig:Figure4}(b) displays many visibilities for the anti-diagonal basis and exemplifies our ability to control the phase term $\gamma$ of Eq. \ref{eq:state:BRW:dichroic}. As alluded to earlier, birefringence was introduced in the idler arm via the insertion of an additional QWP.  By tilting the QWP, the distance the idler photon traveled inside the birefringent material was altered, introducing an additional polarization dependent phase. This changed the polarization correlations between the signal and idler photon.
For a maximally entangled pure state (e.g. $\frac{1}{\sqrt{2}}(|HV\rangle+|VH\rangle)$), the expected behaviour is the conversion of a sinusoid with perfect visibility (minima=0) to an inverted sinusoid and back again, going through a period of flat uncorrelated behaviour when $\gamma=\pi/2$.
That the pair count rates do not approach the noise floor in the experiment for all basis choices is mainly a result of the imperfect overlap between the spectra shown in Fig.~\ref{fig:Figure3}.  Additional arguments for the reduced visibilities are put forth in the discussion.  Nonetheless, the existence of significant interference for all basis choices is consistent with the predictions of Eq.~(\ref{eq:state:BRW:dichroic}), and solidifies the observation of polarization entanglement emerging directly from the BRW.

\begin{figure}
\begin{tabular}{c c}
\subfigure[Real ]{\includegraphics[width=0.45\columnwidth]{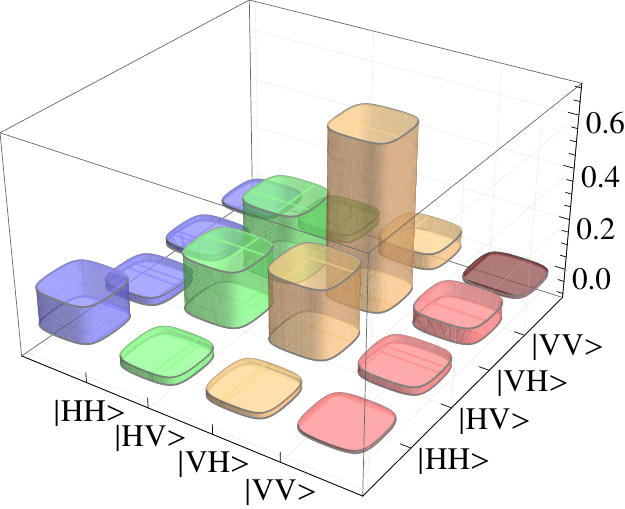}} &
\subfigure[Imaginary ]{\includegraphics[width=0.45\columnwidth]{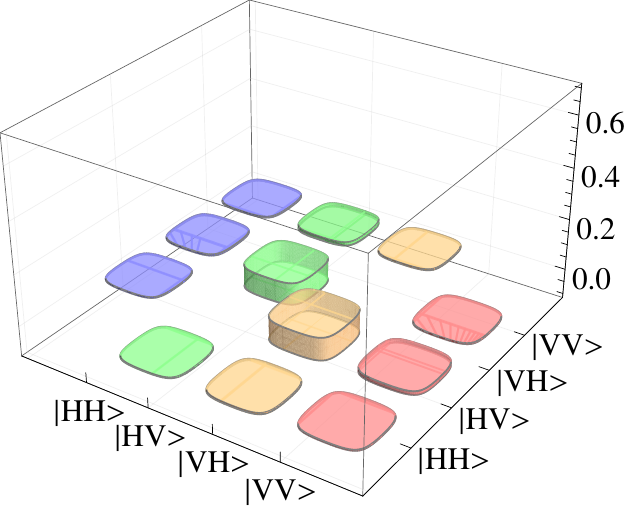}}
\end{tabular}
\caption{Reconstructed polarization state generated by the BRW. The panels show real (a) and imaginary (b) parts of the state. The resulting density matrix entanglement measure, concurrence, equals $0.52$. The fidelity with the Bell state $(\ket{HV}+\ket{VH})/\sqrt{2}$ state is $0.83$ and the purity is $0.64$.}

\label{fig:Figure14}
\end{figure}

To fully quantify the correlations of the photon pair generated by the BRW, quantum state estimation was performed via tomographic measurements \cite{James2001} and the resulting density matrix was reconstructed using the maximum likelihood method \cite{Banaszek2000}. The Methods section explains the measurements in more detail.  The result is shown in Fig.~\ref{fig:Figure14}. Off-diagonal elements are clearly visible, evidence of the non-classical nature of the biphoton state. From the density matrix, various indicators of the entanglement quality can be computed. The concurrence, an entanglement monotone\cite{Hill1997,Nielsen2000}, is found to be $0.52$, while the fidelity \cite{Nielsen2000} with the expected maximally entangled state $(\ket{HV}+\ket{VH})/\sqrt{2}$ is computed to be $0.83$. These measures provide further proof that the BRW is capable of inherently producing polarization entanglement.

\section{Discussion}
The BRW examined here was not intentionally fabricated to produce entangled photons. Designed to optimize pair production and upconversion efficiency \cite{Abolghasem_09-matching1.art}, the spectral overlap is not ideal and degrades the intrinsic entanglement generation capability. An interesting asymmetry observed in this work is the level of the minima measured in the H/V basis. The origins of this small but appreciable H/H background are likely due to residual pump fluorescence and/or polarization hybridization in the V-polarized down converted mode. Nonetheless, these effects are not fundamental limitations, but arise from the current design and fabrication method.  Future samples that are optimized to produce entanglement \cite{Zhukovsky_2012.art} promise to avoid this behaviour.

Despite the many material benefits that GaAs offers (order of magnitude stronger $\chi^{(2)}$ than typical crystals, large transparency window, monolithicity, integratability etc.), it has remained a prized but elusive non-linear material.  Entirely due to its challenging phase matching requirements, GaAs based photon sources have been overshadowed in the quantum community by materials such as, lithium niobate, pottasium titanyl phosphate, and barium borate which are easier to phase match.  Indeed,  in juxtaposition, efforts have been aimed at \emph{integrating} these materials \cite{Sohler_2008.art,Crespi_2011,Sansoni_2012}.  As an alternative to GaAs, high quality integrated quantum photonic circuitry has been developed in silica \cite{Politi_2008,Politi_2009}, but as an entangled photon source, its optical non-linearity is weaker \cite{Matsuda_2012.art}, and still has its challenges to become a fully integrated platform viable for quantum processes.

For GaAs, techniques have been developed to solve the phase matching problem such as form-birefringence \cite{Fiore_96.art}; out of plane pumping \cite{Orieux_2011.art}; or quasi-phase matching by traditional periodic crystal inversion \cite{Schaar_2008.art}.  Further, out of plane pumping has the benefit of producing counter-propagating entangled photons \cite{Orieux2013}.  But while these designs are promising, most are not truly monolithic, and are at greater risk of becoming encumbered with the complications of integration with other photonic components.

In contrast, the BRW design has now shown to be capable of solving all of the above concerns.  It is favourable for becoming its own pump laser \cite{Bijlani_11.art}, it is efficient at producing pairs \cite{Horn_2012.art}, and here we demonstrate a very distinct ``quantum'' opportunity.  Gained by a clever modal phase matching technique that exploits mature nanofabrication technology, the GaAs based BRW can produce useful polarization entangled photons directly and without the need for any post selection, compensation or interferometry. Ironically, its success is tied to the difficulty in phase matching -- GaAs's lack of birefringence turns out to be a virtue for the production of on chip entanglement.  Combined with the many aforementioned benefits that GaAs has to offer, these latest results show that the BRW platform should be seriously considered among the integrated quantum photonics community as one of the most promising platforms on which to build integrated optical quantum information processing devices.

\section{Acknowledgements}
The authors  acknowledge  funding from NSERC (CGS, QuantumWorks, Discovery, USRA), Ontario Ministry of Research and Innovation (ERA program and research infrastructure program), CIFAR, Industry Canada, CFI, and CMC Microsystems. P. Kolenderski acknowledges support from the Mobility Plus project financed by the Polish Ministry of Science and Higher Education.

\section{Methods}
\textbf{Device description and experimental setup}
The $3.8~\mu$m wide, $2.2$~mm long ridge BRW was fabricated via metal-organic chemical vapour deposition.  Details of the epitaxy can be found in Ref. \cite{Abolghasem_2012.art}, where it is referred to as BRW$_1$.  The waveguide was placed into an objectively coupled ``end-fire rig'' \cite{Horn_2012.art} and pumped with horizontally polarized light at a wavelength $\lambda_p$, and power, $P$.  The light was focussed onto the front facet and directed into the core region of the BRW.  Output light emerging from the core at the back facet was filtered to remove the pump before being collected for measurement.

For the single photon spectrum measurements, $P$ was approximately 20-30~mW.  The output was transmitted through a polarizing beam splitter (PBS) separating the light into distinct H and V ports. Individually, both output ports were sent into a Czerny-Turner type monochromator equipped with a novel, high timing resolution, free space InGaAs/InP single photon avalanche diode \cite{Tosi2012a}. Spectra for both polarizations were obtained for the range of CW-pump central frequencies from $\lambda_p=776-778$~nm in increments of about 0.1-0.2~nm.

To characterize polarization entanglement, $P$ was set to approximately $1.2$~mW, and the PBS was replaced by a DBS with a central wavelength $\lambda\approx$ 1560~nm, and a width of $\approx$ 10~nm.  The pump wavelength was set to $\lambda=777.9$~nm, which produced photon pairs via SPDC with wavelengths near $\lambda=1537$ (signal)~nm and $\lambda=1575$~nm (idler). Lower energy ($\lambda_i>$1560~nm) idler photons were transmitted, while higher energy ($\lambda_s<$1560~nm) signal photons were reflected.  A polarization analyzer, consisting of a QWP, a half wave plate (HWP), and a linear sheet polarizer, was placed in each of the signal and idler arms. An additional QWP, which could be tilted, was placed in the idler arm to control the relative phase $\gamma$ between HV and VH pairs. Each arm collected photons into a multi-mode fiber connected to a single photon detector (id201, idQuantique).  The signal arm detector was internally triggered at $1$ MHz, open for $20$~ns and set for $10$\% quantum efficiency.  Count rates were approximately $7000 \pm 500$ counts per second (cps).  The idler arm -- gated by counts recorded in the signal arm -- was set at 15\% efficiency and the gate time was set to $5$~ns.  Optical and electronic delays were adjusted so that the arrival of the idler photon coincided with the electronic gate signal.  Thus, idler count rates were effectively pair count rates, and were anywhere from 7-100 cps.

For the tomographic reconstruction of the state, an overcomplete set of 36 polarization measurements were performed; setting the polarization analyzers for every combination of H,V,anti-diagonal, diagonal, left-circular, and right-circular. Each measurement had a duration of 2 minutes.  Background count rates were obtained by changing the electronic delay in the idler arm to ensure the observation of random background events. \\


\begin{thebibliography}{28}
\expandafter\ifx\csname natexlab\endcsname\relax\def\natexlab#1{#1}\fi
\expandafter\ifx\csname bibnamefont\endcsname\relax
  \def\bibnamefont#1{#1}\fi
\expandafter\ifx\csname bibfnamefont\endcsname\relax
  \def\bibfnamefont#1{#1}\fi
\expandafter\ifx\csname citenamefont\endcsname\relax
  \def\citenamefont#1{#1}\fi
\expandafter\ifx\csname url\endcsname\relax
  \def\url#1{\texttt{#1}}\fi
\expandafter\ifx\csname urlprefix\endcsname\relax\def\urlprefix{URL }\fi
\providecommand{\bibinfo}[2]{#2}
\providecommand{\eprint}[2][]{\url{#2}}

\bibitem[{\citenamefont{Louisell et~al.}(1961)\citenamefont{Louisell, Yariv,
  and Siegman}}]{Louisell_1961.art}
\bibinfo{author}{\bibfnamefont{W.~H.} \bibnamefont{Louisell}},
  \bibinfo{author}{\bibfnamefont{A.}~\bibnamefont{Yariv}}, \bibnamefont{and}
  \bibinfo{author}{\bibfnamefont{A.~E.} \bibnamefont{Siegman}},
  \bibinfo{journal}{Phys. Rev.} \textbf{\bibinfo{volume}{124}},
  \bibinfo{pages}{1646} (\bibinfo{year}{1961}).

\bibitem[{\citenamefont{{Klyshko}}(1967)}]{Klyshko_1967.art}
\bibinfo{author}{\bibfnamefont{D.~N.} \bibnamefont{{Klyshko}}},
  \bibinfo{journal}{Soviet Journal of Experimental and Theoretical Physics
  Letters} \textbf{\bibinfo{volume}{6}}, \bibinfo{pages}{23}
  (\bibinfo{year}{1967}).

\bibitem[{\citenamefont{Kim and Grice}(2002)}]{Kim2002}
\bibinfo{author}{\bibfnamefont{Y.-H.} \bibnamefont{Kim}} \bibnamefont{and}
  \bibinfo{author}{\bibfnamefont{W.~P.} \bibnamefont{Grice}},
  \bibinfo{journal}{J. Mod. Opt.} \textbf{\bibinfo{volume}{49}},
  \bibinfo{pages}{2309} (\bibinfo{year}{2002}), \eprint{quant-ph/0207047}.

\bibitem[{\citenamefont{Ueno et~al.}(2012)\citenamefont{Ueno, {Kaneda},
  {Suzuki}, {Nagano}, {Syouji}, {Shimizu}, {Suizu}, and {Edamatsu}}}]{Ueno2012}
\bibinfo{author}{\bibfnamefont{W.}~\bibnamefont{Ueno}},
  \bibinfo{author}{\bibfnamefont{F.}~\bibnamefont{{Kaneda}}},
  \bibinfo{author}{\bibfnamefont{H.}~\bibnamefont{{Suzuki}}},
  \bibinfo{author}{\bibfnamefont{S.}~\bibnamefont{{Nagano}}},
  \bibinfo{author}{\bibfnamefont{A.}~\bibnamefont{{Syouji}}},
  \bibinfo{author}{\bibfnamefont{R.}~\bibnamefont{{Shimizu}}},
  \bibinfo{author}{\bibfnamefont{K.}~\bibnamefont{{Suizu}}}, \bibnamefont{and}
  \bibinfo{author}{\bibfnamefont{K.}~\bibnamefont{{Edamatsu}}},
  \bibinfo{journal}{Opt. Express} \textbf{\bibinfo{volume}{20}},
  \bibinfo{pages}{5508} (\bibinfo{year}{2012}).

\bibitem[{\citenamefont{Yao et~al.}(2012)\citenamefont{Yao, Wang, Xu, Lu, Pan,
  Bao, Peng, Lu, Chen, and Pan}}]{Yao_2012.art}
\bibinfo{author}{\bibfnamefont{X.-C.} \bibnamefont{Yao}},
  \bibinfo{author}{\bibfnamefont{T.-X.} \bibnamefont{Wang}},
  \bibinfo{author}{\bibfnamefont{P.}~\bibnamefont{Xu}},
  \bibinfo{author}{\bibfnamefont{H.}~\bibnamefont{Lu}},
  \bibinfo{author}{\bibfnamefont{G.-S.} \bibnamefont{Pan}},
  \bibinfo{author}{\bibfnamefont{X.-H.} \bibnamefont{Bao}},
  \bibinfo{author}{\bibfnamefont{C.-Z.} \bibnamefont{Peng}},
  \bibinfo{author}{\bibfnamefont{C.-Y.} \bibnamefont{Lu}},
  \bibinfo{author}{\bibfnamefont{Y.-A.} \bibnamefont{Chen}}, \bibnamefont{and}
  \bibinfo{author}{\bibfnamefont{J.-W.} \bibnamefont{Pan}},
  \bibinfo{journal}{Nature Photon.} \textbf{\bibinfo{volume}{6}},
  \bibinfo{pages}{225} (\bibinfo{year}{2012}).

\bibitem[{\citenamefont{O'Brien}(2007)}]{O'Brien_2007}
\bibinfo{author}{\bibfnamefont{J.~L.} \bibnamefont{O'Brien}},
  \bibinfo{journal}{Science} \textbf{\bibinfo{volume}{318}},
  \bibinfo{pages}{1567} (\bibinfo{year}{2007}).

\bibitem[{\citenamefont{Politi et~al.}(2008)\citenamefont{Politi, Cryan,
  Rarity, Yu, and O'Brien}}]{Politi_2008}
\bibinfo{author}{\bibfnamefont{A.}~\bibnamefont{Politi}},
  \bibinfo{author}{\bibfnamefont{M.~J.} \bibnamefont{Cryan}},
  \bibinfo{author}{\bibfnamefont{J.~G.} \bibnamefont{Rarity}},
  \bibinfo{author}{\bibfnamefont{S.}~\bibnamefont{Yu}}, \bibnamefont{and}
  \bibinfo{author}{\bibfnamefont{J.~L.} \bibnamefont{O'Brien}},
  \bibinfo{journal}{Science} \textbf{\bibinfo{volume}{320}},
  \bibinfo{pages}{646} (\bibinfo{year}{2008}).

\bibitem[{\citenamefont{Horn et~al.}(2012)\citenamefont{Horn, Abolghasem,
  Bijlani, Kang, Helmy, and Weihs}}]{Horn_2012.art}
\bibinfo{author}{\bibfnamefont{R.}~\bibnamefont{Horn}},
  \bibinfo{author}{\bibfnamefont{P.}~\bibnamefont{Abolghasem}},
  \bibinfo{author}{\bibfnamefont{B.~J.} \bibnamefont{Bijlani}},
  \bibinfo{author}{\bibfnamefont{D.}~\bibnamefont{Kang}},
  \bibinfo{author}{\bibfnamefont{A.~S.} \bibnamefont{Helmy}}, \bibnamefont{and}
  \bibinfo{author}{\bibfnamefont{G.}~\bibnamefont{Weihs}},
  \bibinfo{journal}{Phys. Rev. Lett.} \textbf{\bibinfo{volume}{108}},
  \bibinfo{pages}{153605} (\bibinfo{year}{2012}).

\bibitem[{\citenamefont{Valles et~al.}(2013)\citenamefont{Valles, Hendrych,
  Svozilik, Machulka, Abolghasem, Kang, Bijlani, Helmy, and
  Torres}}]{Valles2013}
\bibinfo{author}{\bibfnamefont{A.}~\bibnamefont{Valles}},
  \bibinfo{author}{\bibfnamefont{M.}~\bibnamefont{Hendrych}},
  \bibinfo{author}{\bibfnamefont{J.}~\bibnamefont{Svozilik}},
  \bibinfo{author}{\bibfnamefont{R.}~\bibnamefont{Machulka}},
  \bibinfo{author}{\bibfnamefont{P.}~\bibnamefont{Abolghasem}},
  \bibinfo{author}{\bibfnamefont{D.}~\bibnamefont{Kang}},
  \bibinfo{author}{\bibfnamefont{B.~J.} \bibnamefont{Bijlani}},
  \bibinfo{author}{\bibfnamefont{A.~S.} \bibnamefont{Helmy}}, \bibnamefont{and}
  \bibinfo{author}{\bibfnamefont{J.~P.} \bibnamefont{Torres}}
  (\bibinfo{year}{2013}), \eprint{arXiv:1303.3406}.

\bibitem[{\citenamefont{Bijlani et~al.}(2011)\citenamefont{Bijlani, Abolghasem,
  and Helmy}}]{Bijlani_11.art}
\bibinfo{author}{\bibfnamefont{B.~J.} \bibnamefont{Bijlani}},
  \bibinfo{author}{\bibfnamefont{P.}~\bibnamefont{Abolghasem}},
  \bibnamefont{and} \bibinfo{author}{\bibfnamefont{A.~S.} \bibnamefont{Helmy}},
  in \emph{\bibinfo{booktitle}{CLEO:2011 - Laser Applications to Photonic
  Applications}} (\bibinfo{publisher}{Optical Society of America},
  \bibinfo{year}{2011}), p. \bibinfo{pages}{PDPA3}.

\bibitem[{\citenamefont{Helmy}(2006)}]{Helmy_06.art}
\bibinfo{author}{\bibfnamefont{A.~S.} \bibnamefont{Helmy}},
  \bibinfo{journal}{Opt. Express} \textbf{\bibinfo{volume}{14}},
  \bibinfo{pages}{1243} (\bibinfo{year}{2006}).

\bibitem[{\citenamefont{James et~al.}(2001)\citenamefont{James, {Kwiat},
  {Munro}, and {White}}}]{James2001}
\bibinfo{author}{\bibfnamefont{D.~F.~V.} \bibnamefont{James}},
  \bibinfo{author}{\bibfnamefont{P.~G.} \bibnamefont{{Kwiat}}},
  \bibinfo{author}{\bibfnamefont{W.~J.} \bibnamefont{{Munro}}},
  \bibnamefont{and} \bibinfo{author}{\bibfnamefont{A.~G.}
  \bibnamefont{{White}}}, \bibinfo{journal}{Phys. Rev. A}
  \textbf{\bibinfo{volume}{64}}, \bibinfo{pages}{052312}
  (\bibinfo{year}{2001}), \eprint{arXiv:quant-ph/0103121}.

\bibitem[{\citenamefont{Banaszek et~al.}(2000)\citenamefont{Banaszek,
  {D'ariano}, {Paris}, and {Sacchi}}}]{Banaszek2000}
\bibinfo{author}{\bibfnamefont{K.}~\bibnamefont{Banaszek}},
  \bibinfo{author}{\bibfnamefont{G.~M.} \bibnamefont{{D'ariano}}},
  \bibinfo{author}{\bibfnamefont{M.~G.} \bibnamefont{{Paris}}},
  \bibnamefont{and} \bibinfo{author}{\bibfnamefont{M.~F.}
  \bibnamefont{{Sacchi}}}, \bibinfo{journal}{Phys. Rev. A}
  \textbf{\bibinfo{volume}{61}}, \bibinfo{pages}{010304}
  (\bibinfo{year}{2000}), \eprint{arXiv:quant-ph/9909052}.

\bibitem[{\citenamefont{Hill and Wootters}(1997)}]{Hill1997}
\bibinfo{author}{\bibfnamefont{S.}~\bibnamefont{Hill}} \bibnamefont{and}
  \bibinfo{author}{\bibfnamefont{W.~K.} \bibnamefont{Wootters}},
  \bibinfo{journal}{Phys. Rev. Lett.} \textbf{\bibinfo{volume}{78}},
  \bibinfo{pages}{5022} (\bibinfo{year}{1997}),
  \eprint{arXiv:quant-ph/9703041}.

\bibitem[{\citenamefont{Nielsen and Chuang}(2000)}]{Nielsen2000}
\bibinfo{author}{\bibfnamefont{M.~A.} \bibnamefont{Nielsen}} \bibnamefont{and}
  \bibinfo{author}{\bibfnamefont{I.~L.} \bibnamefont{Chuang}},
  \emph{\bibinfo{title}{Quantum Computation and Quantum Information}}
  (\bibinfo{publisher}{Cambridge University Press}, \bibinfo{year}{2000}),
  \bibinfo{edition}{1st} ed.

\bibitem[{\citenamefont{Abolghasem and
  Helmy}(2009)}]{Abolghasem_09-matching1.art}
\bibinfo{author}{\bibfnamefont{P.}~\bibnamefont{Abolghasem}} \bibnamefont{and}
  \bibinfo{author}{\bibfnamefont{A.}~\bibnamefont{Helmy}},
  \bibinfo{journal}{IEEE J. Quant. Electron.} \textbf{\bibinfo{volume}{45}},
  \bibinfo{pages}{646 } (\bibinfo{year}{2009}).

\bibitem[{\citenamefont{Zhukovsky et~al.}(2012)\citenamefont{Zhukovsky, Helt,
  Kang, Abolghasem, Helmy, and Sipe}}]{Zhukovsky_2012.art}
\bibinfo{author}{\bibfnamefont{S.~V.} \bibnamefont{Zhukovsky}},
  \bibinfo{author}{\bibfnamefont{L.~G.} \bibnamefont{Helt}},
  \bibinfo{author}{\bibfnamefont{D.}~\bibnamefont{Kang}},
  \bibinfo{author}{\bibfnamefont{P.}~\bibnamefont{Abolghasem}},
  \bibinfo{author}{\bibfnamefont{A.~S.} \bibnamefont{Helmy}}, \bibnamefont{and}
  \bibinfo{author}{\bibfnamefont{J.~E.} \bibnamefont{Sipe}},
  \bibinfo{journal}{Phys. Rev. A} \textbf{\bibinfo{volume}{85}},
  \bibinfo{eid}{013838} (\bibinfo{year}{2012}).

\bibitem[{\citenamefont{Sohler et~al.}(2008)\citenamefont{Sohler, Hu, Ricken,
  Quiring, Vannahme, Herrmann, B\"{u}chter, Reza, Grundk\"{o}tter, Orlov
  et~al.}}]{Sohler_2008.art}
\bibinfo{author}{\bibfnamefont{W.}~\bibnamefont{Sohler}},
  \bibinfo{author}{\bibfnamefont{H.}~\bibnamefont{Hu}},
  \bibinfo{author}{\bibfnamefont{R.}~\bibnamefont{Ricken}},
  \bibinfo{author}{\bibfnamefont{V.}~\bibnamefont{Quiring}},
  \bibinfo{author}{\bibfnamefont{C.}~\bibnamefont{Vannahme}},
  \bibinfo{author}{\bibfnamefont{H.}~\bibnamefont{Herrmann}},
  \bibinfo{author}{\bibfnamefont{D.}~\bibnamefont{B\"{u}chter}},
  \bibinfo{author}{\bibfnamefont{S.}~\bibnamefont{Reza}},
  \bibinfo{author}{\bibfnamefont{W.}~\bibnamefont{Grundk\"{o}tter}},
  \bibinfo{author}{\bibfnamefont{S.}~\bibnamefont{Orlov}},
  \bibnamefont{et~al.}, \bibinfo{journal}{Opt. Photon. News}
  \textbf{\bibinfo{volume}{19}}, \bibinfo{pages}{24} (\bibinfo{year}{2008}).

\bibitem[{\citenamefont{Crespi et~al.}(2011)\citenamefont{Crespi, Ramponi,
  Osellame, Sansoni, Bongioanni, Sciarrino, Vallone, and
  Mataloni}}]{Crespi_2011}
\bibinfo{author}{\bibfnamefont{A.}~\bibnamefont{Crespi}},
  \bibinfo{author}{\bibfnamefont{R.}~\bibnamefont{Ramponi}},
  \bibinfo{author}{\bibfnamefont{R.}~\bibnamefont{Osellame}},
  \bibinfo{author}{\bibfnamefont{L.}~\bibnamefont{Sansoni}},
  \bibinfo{author}{\bibfnamefont{I.}~\bibnamefont{Bongioanni}},
  \bibinfo{author}{\bibfnamefont{F.}~\bibnamefont{Sciarrino}},
  \bibinfo{author}{\bibfnamefont{G.}~\bibnamefont{Vallone}}, \bibnamefont{and}
  \bibinfo{author}{\bibfnamefont{P.}~\bibnamefont{Mataloni}},
  \bibinfo{journal}{Nat. Commun.} \textbf{\bibinfo{volume}{2}}
  (\bibinfo{year}{2011}).

\bibitem[{\citenamefont{Sansoni et~al.}(2012)\citenamefont{Sansoni, Sciarrino,
  Vallone, Mataloni, Crespi, Ramponi, and Osellame}}]{Sansoni_2012}
\bibinfo{author}{\bibfnamefont{L.}~\bibnamefont{Sansoni}},
  \bibinfo{author}{\bibfnamefont{F.}~\bibnamefont{Sciarrino}},
  \bibinfo{author}{\bibfnamefont{G.}~\bibnamefont{Vallone}},
  \bibinfo{author}{\bibfnamefont{P.}~\bibnamefont{Mataloni}},
  \bibinfo{author}{\bibfnamefont{A.}~\bibnamefont{Crespi}},
  \bibinfo{author}{\bibfnamefont{R.}~\bibnamefont{Ramponi}}, \bibnamefont{and}
  \bibinfo{author}{\bibfnamefont{R.}~\bibnamefont{Osellame}},
  \bibinfo{journal}{Phys. Rev. Lett.} \textbf{\bibinfo{volume}{108}},
  \bibinfo{pages}{010502} (\bibinfo{year}{2012}).

\bibitem[{\citenamefont{Politi et~al.}(2009)\citenamefont{Politi, Matthews, and
  O'Brien}}]{Politi_2009}
\bibinfo{author}{\bibfnamefont{A.}~\bibnamefont{Politi}},
  \bibinfo{author}{\bibfnamefont{J.~C.~F.} \bibnamefont{Matthews}},
  \bibnamefont{and} \bibinfo{author}{\bibfnamefont{J.~L.}
  \bibnamefont{O'Brien}}, \bibinfo{journal}{Science}
  \textbf{\bibinfo{volume}{325}}, \bibinfo{pages}{1221} (\bibinfo{year}{2009}).

\bibitem[{\citenamefont{Matsuda et~al.}(2012)\citenamefont{Matsuda, Le~Jeannic,
  Fukuda, Tsuchizawa, Munro, Shimizu, Yamada, Tokura, and
  Takesue}}]{Matsuda_2012.art}
\bibinfo{author}{\bibfnamefont{N.}~\bibnamefont{Matsuda}},
  \bibinfo{author}{\bibfnamefont{H.}~\bibnamefont{Le~Jeannic}},
  \bibinfo{author}{\bibfnamefont{H.}~\bibnamefont{Fukuda}},
  \bibinfo{author}{\bibfnamefont{T.}~\bibnamefont{Tsuchizawa}},
  \bibinfo{author}{\bibfnamefont{W.~J.} \bibnamefont{Munro}},
  \bibinfo{author}{\bibfnamefont{K.}~\bibnamefont{Shimizu}},
  \bibinfo{author}{\bibfnamefont{K.}~\bibnamefont{Yamada}},
  \bibinfo{author}{\bibfnamefont{Y.}~\bibnamefont{Tokura}}, \bibnamefont{and}
  \bibinfo{author}{\bibfnamefont{H.}~\bibnamefont{Takesue}},
  \bibinfo{journal}{Sci. Rep.} \textbf{\bibinfo{volume}{2}},
  \bibinfo{pages}{812} (\bibinfo{year}{2012}).

\bibitem[{\citenamefont{Fiore et~al.}(1996)\citenamefont{Fiore, Rosencher,
  Berger, Laurent, Vodjdani, and Nagle}}]{Fiore_96.art}
\bibinfo{author}{\bibfnamefont{A.}~\bibnamefont{Fiore}},
  \bibinfo{author}{\bibfnamefont{E.}~\bibnamefont{Rosencher}},
  \bibinfo{author}{\bibfnamefont{V.}~\bibnamefont{Berger}},
  \bibinfo{author}{\bibfnamefont{N.}~\bibnamefont{Laurent}},
  \bibinfo{author}{\bibfnamefont{N.}~\bibnamefont{Vodjdani}}, \bibnamefont{and}
  \bibinfo{author}{\bibfnamefont{J.}~\bibnamefont{Nagle}}, in
  \emph{\bibinfo{booktitle}{Conference on Lasers and Electro-Optics (CLEO)
  1996}} (\bibinfo{year}{1996}), pp. \bibinfo{pages}{231 -- 232}.

\bibitem[{\citenamefont{Orieux et~al.}(2011)\citenamefont{Orieux, Caillet,
  Lema\^{i}tre, Filloux, Favero, Leo, and Ducci}}]{Orieux_2011.art}
\bibinfo{author}{\bibfnamefont{A.}~\bibnamefont{Orieux}},
  \bibinfo{author}{\bibfnamefont{X.}~\bibnamefont{Caillet}},
  \bibinfo{author}{\bibfnamefont{A.}~\bibnamefont{Lema\^{i}tre}},
  \bibinfo{author}{\bibfnamefont{P.}~\bibnamefont{Filloux}},
  \bibinfo{author}{\bibfnamefont{I.}~\bibnamefont{Favero}},
  \bibinfo{author}{\bibfnamefont{G.}~\bibnamefont{Leo}}, \bibnamefont{and}
  \bibinfo{author}{\bibfnamefont{S.}~\bibnamefont{Ducci}}, \bibinfo{journal}{J.
  Opt. Soc. Am. B} \textbf{\bibinfo{volume}{28}}, \bibinfo{pages}{45}
  (\bibinfo{year}{2011}).

\bibitem[{\citenamefont{Schaar et~al.}(2008)\citenamefont{Schaar, Vodopyanov,
  Kuo, Fejer, Lin, Yu, Harris, Bliss, Lynch, Kozlov et~al.}}]{Schaar_2008.art}
\bibinfo{author}{\bibfnamefont{J.}~\bibnamefont{Schaar}},
  \bibinfo{author}{\bibfnamefont{K.}~\bibnamefont{Vodopyanov}},
  \bibinfo{author}{\bibfnamefont{P.}~\bibnamefont{Kuo}},
  \bibinfo{author}{\bibfnamefont{M.}~\bibnamefont{Fejer}},
  \bibinfo{author}{\bibfnamefont{A.}~\bibnamefont{Lin}},
  \bibinfo{author}{\bibfnamefont{X.}~\bibnamefont{Yu}},
  \bibinfo{author}{\bibfnamefont{J.}~\bibnamefont{Harris}},
  \bibinfo{author}{\bibfnamefont{D.}~\bibnamefont{Bliss}},
  \bibinfo{author}{\bibfnamefont{C.}~\bibnamefont{Lynch}},
  \bibinfo{author}{\bibfnamefont{V.}~\bibnamefont{Kozlov}},
  \bibnamefont{et~al.}, in \emph{\bibinfo{booktitle}{Lasers and Electro-Optics,
  2008 and 2008 Conference on Quantum Electronics and Laser Science. CLEO/QELS
  2008. Conference on}} (\bibinfo{year}{2008}), pp. \bibinfo{pages}{1 --2}.

\bibitem[{\citenamefont{Orieux et~al.}(2013)\citenamefont{Orieux, Eckstein,
  Lemaître, Filloux, Favero, Leo, Coudreau, Keller, Milman, and
  Ducci}}]{Orieux2013}
\bibinfo{author}{\bibfnamefont{A.}~\bibnamefont{Orieux}},
  \bibinfo{author}{\bibfnamefont{A.}~\bibnamefont{Eckstein}},
  \bibinfo{author}{\bibfnamefont{A.}~\bibnamefont{Lemaître}},
  \bibinfo{author}{\bibfnamefont{P.}~\bibnamefont{Filloux}},
  \bibinfo{author}{\bibfnamefont{I.}~\bibnamefont{Favero}},
  \bibinfo{author}{\bibfnamefont{G.}~\bibnamefont{Leo}},
  \bibinfo{author}{\bibfnamefont{T.}~\bibnamefont{Coudreau}},
  \bibinfo{author}{\bibfnamefont{A.}~\bibnamefont{Keller}},
  \bibinfo{author}{\bibfnamefont{P.}~\bibnamefont{Milman}}, \bibnamefont{and}
  \bibinfo{author}{\bibfnamefont{S.}~\bibnamefont{Ducci}}
  (\bibinfo{year}{2013}), \eprint{arXiv:1301.1764}.

\bibitem[{\citenamefont{Abolghasem et~al.}({2012})\citenamefont{Abolghasem,
  Han, Kang, Bijlani, and Helmy}}]{Abolghasem_2012.art}
\bibinfo{author}{\bibfnamefont{P.}~\bibnamefont{Abolghasem}},
  \bibinfo{author}{\bibfnamefont{J.-B.} \bibnamefont{Han}},
  \bibinfo{author}{\bibfnamefont{D.}~\bibnamefont{Kang}},
  \bibinfo{author}{\bibfnamefont{B.~J.} \bibnamefont{Bijlani}},
  \bibnamefont{and} \bibinfo{author}{\bibfnamefont{A.~S.} \bibnamefont{Helmy}},
  \bibinfo{journal}{IEEE J. Sel. Topics in Quantum Electron.}
  \textbf{\bibinfo{volume}{{18}}}, \bibinfo{pages}{812}
  (\bibinfo{year}{{2012}}).

\bibitem[{\citenamefont{Tosi et~al.}(2012)\citenamefont{Tosi, {Della Frera},
  {Bahgat Shehata}, and {Scarcella}}}]{Tosi2012a}
\bibinfo{author}{\bibfnamefont{A.}~\bibnamefont{Tosi}},
  \bibinfo{author}{\bibfnamefont{A.}~\bibnamefont{{Della Frera}}},
  \bibinfo{author}{\bibfnamefont{A.}~\bibnamefont{{Bahgat Shehata}}},
  \bibnamefont{and}
  \bibinfo{author}{\bibfnamefont{C.}~\bibnamefont{{Scarcella}}},
  \bibinfo{journal}{Rev. Sci. Instrum.} \textbf{\bibinfo{volume}{83}},
  \bibinfo{pages}{013104} (\bibinfo{year}{2012}).

\end{thebibliography}

\newpage
\appendix
\section{Polarization entanglement analysis}
We consider a situation where the BRW is CW-pumped at $\lambda_p=777.9$ nm and where pairs produced at this pump wavelength are pre-dominantly wavelength \emph{non-degenerate}, $\lambda_s\neq\lambda_i$.

\begin{figure}[h]
\includegraphics[width=0.9\columnwidth]{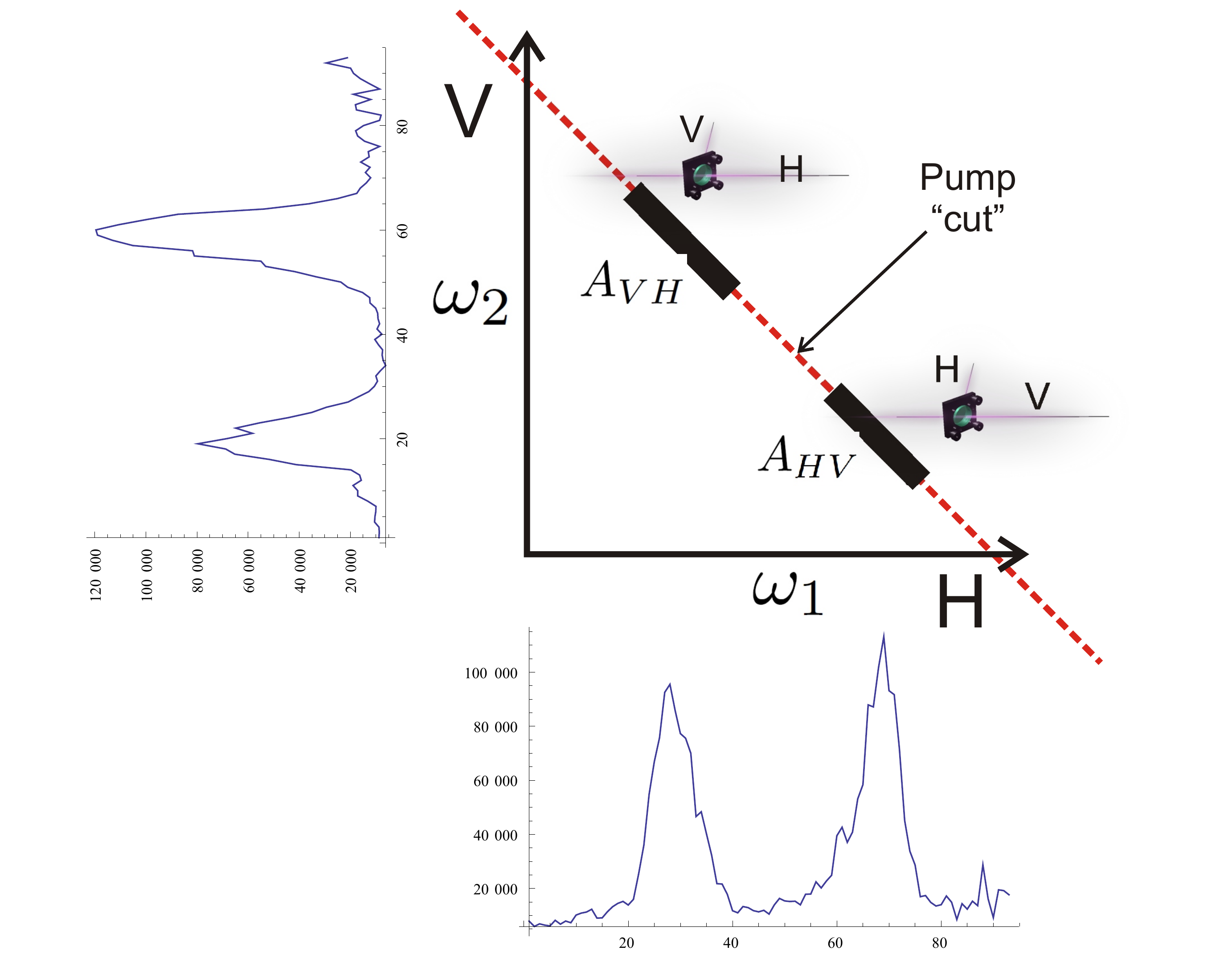}
\caption{Phase matching sketch.  Horizontally polarized photons with frequency $\omega_1$ along the abscissa; Vertically polarized photons with frequency $\omega_2$ along the ordinate axis.  The pump frequency (dashed line) dictates what photons are allowed to decay (black regions).  The marginal spectrums are plotted to the side of both co-ordinate axis and are examples of a vertical slice through both spectra observed in Fig.~\ref{fig:Figure3}.  Insets show the DBS, indicating what it does to the photons created in each region selected by the pump.}
\label{fig:Figure_supplementary_aid}
\end{figure}

Fig.~\ref{fig:Figure_supplementary_aid} sketches the expected photon pair spectral amplitude in standard frequency co-ordinates with H polarized photons of frequency $\omega_1$ along the abscissa and V polarized photons with frequency $\omega_2$ along the ordinate axis. The unusual shape of the biphoton spectral amplitude allows us to approximate the resulting BRW output state in the following way:
\begin{multline}
\ket{\Psi}_\text{BRW}=\\
\int d\omega_1 d\omega_2\left[A_{HV}(\omega_1,\omega_2)+A_{VH}(\omega_1,\omega_2)\right]\ket{\omega_1}_H\ket{\omega_2}_V,
\label{eq:state:BRW}
\end{multline}
where $\ket{\omega}_\sigma$ denotes a $\sigma$-polarized photon at frequency $\omega$, $\sigma\in\{H,V\}$. The total spectral amplitude is proportional to the sum of the two amplitudes, $A_{HV}(\omega_1,\omega_2)$ + $A_{VH}(\omega_1,\omega_2)$. The amplitude $A_{HV}(\omega_1,\omega_2)$ governs the pair where the H photon has the shorter wavelength; $A_{VH}(\omega_1,\omega_2)$ governs the pair where the V polarized photon has the shorter wavelength.  The measured spectra, presented in Fig.~\ref{fig:Figure3} are proportional to the marginals of the sum of the two amplitudes.  Specifically, as a function of the pump frequency, the surface plot in Fig.~\ref{fig:Figure3}(a) is proportional to $\int d\omega_1 |A_{HV}(\omega_1,\omega_2)+A_{VH}(\omega_1,\omega_2)|^2$ and the surface plot in Fig.~\ref{fig:Figure3}(b) is proportional to $\int d\omega_2 |A_{HV}(\omega_1,\omega_2)+A_{VH}(\omega_1,\omega_2)|^2$.

We now consider the biphoton state under the action of the DBS with a center frequency $\omega_\text{DBS}\approx$1560~nm, and which transmits `idler' photons with longer wavelengths, and reflects `signal' photons with shorter wavelengths. Referring again to Fig.~\ref{fig:Figure_supplementary_aid}, consider the pair $\ket{\omega_1,H}_s\ket{\omega_2,V}_i$, where the H (V) polarized photon at frequency $\omega_1>\omega_\text{DBS}$ ($\omega_2<\omega_\text{DBS}$) takes the signal (idler) output port of the DBS . The spectral amplitude of this pair is given by $A_{HV}(\omega_1,\omega_2)$. Analogously, the spectral amplitude of a pair $\ket{\omega_2,V}_s\ket{\omega_1,H}$, where $\omega_1<\omega_\text{DBS}$ and $\omega_2>\omega_\text{DBS}$ is given by $A_{VH}(\omega_1,\omega_2)$. Note that we have judiciously chosen the pump frequency so that the bulk of the photon frequencies involved here are far enough away from the DBS central frequency $\omega_\text{DBS}$. Thus we can safely ignore events where both photons are routed to the same output port of the DBS.  This assumption also implies that the effect of the dichroic on the amplitudes $A_{HV}(\omega_1,\omega_2)$ and $A_{VH}(\omega_1,\omega_2)$ are minimal, and so we retain the same labeling before and after the DBS transformation.  The resulting state can be written as in Eq.~\ref{eq:state:BRW:dichroic}:
\begin{multline*}
\ket{\Psi}=\int d\omega_1 d\omega_2 \left(A_{HV}(\omega_1,\omega_2)\ket{\omega_1,H}_s\ket{\omega_2,V}_i+\right. \\\left.A_{VH}(\omega_1,\omega_2)\ket{\omega_2,V}_s\ket{\omega_1,H}_i\right),
\end{multline*}
where $\ket{\omega,\sigma}_\mu$, $\sigma\in\{H,V\}$ and $\mu\in\{s,i\}$, denotes a $\sigma$-polarized photon at frequency $\omega$ in the $\mu$ output port of the DBS. Note the index interchange in the second term which arises from the transformation of the DBS.
Due to design related asymmetries in the phase matching function, the amplitude $A_{HV}(\omega_1,\omega_2)$ does not equal the amplitude $A_{VH}(\omega_2,\omega_1)$ for all $(\omega_1,\omega_2)$. This is the main reason for the reduced visibilities and entanglement quality observed in the experiment.

Finally, we analyze the resulting polarization state of the photon pair when the spectral information is disregarded. This produces a density matrix in the following form:
\begin{multline}
\hat{\rho} = p_{HV} \ket{HV}\bra{HV}+p_{VH} \ket{VH}\bra{VH}\\+q\ket{HV}\bra{VH}+q^*\ket{VH}\bra{HV}
\label{eq:dm}
\end{multline}
where the coefficients $p_{HV}$, $p_{VH}$ and $q$ are related to the spectra of the generated photons as:
\begin{eqnarray}
p_{HV}&=&\int d\omega_1 d\omega_2 |A_{HV}(\omega_1,\omega_2)|^2,\\
p_{VH}&=&\int d\omega_1 d\omega_2 |A_{VH}(\omega_1,\omega_2)|^2,\\
q&=&\int d\omega_1 d\omega_2 A_{HV}(\omega_1,\omega_2)A_{VH}^*(\omega_2,\omega_1)e^{i \gamma}.
\end{eqnarray}
Note that we explicitly insert a phase term $\gamma$, as it is experimentally accessible, and can be controlled by introducing additional birefringence in one of the paths of the biphoton. It is also worth emphasizing that if the overlap is perfect then $|q|=1/2$ and the state produced by the BRW would be maximally entangled in the polarization degree of freedom.

The above analysis allows us to predict the entanglement visibilities based upon the observed spectrum of the generated photons.  As in the experiment, when the signal polarizer is set to transmit anti-diagonal polarization, and the idler polarizer is rotated about an axis perpendicular to its face by some angle $\theta$, then the observed coincidence count, C, is proportional to
$(2 \text{Re} (q) {\sin}(2 \theta))/2$.  It is now clear that the visibility, defined as $[\max_{\theta}(C)-\min_{\theta}(C)]/[\max_{\theta}(C)+\min_{\theta}(C)]$ equals  $2q$. For the measured spectra, optimal overlap occurs when pumping at a central wavelength of $777.9$ nm, where based on the measured data we estimate  $2q\approx 0.8$.

\end{document}